\documentclass[aps,prd,preprint,groupedaddress,nofootinbib,longbibliography]{revtex4-1}

\usepackage{amssymb}
\usepackage{amsmath}
\usepackage{hyperref}
\usepackage{slashed}
\usepackage[utf8]{inputenc}

\begin{document}

\title{Remarks on a Gauge Theory for Continuous Spin Particles}

\author{Victor O. Rivelles}
\affiliation{ Instituto de F\'{i}sica, Universidade de S\~ao Paulo,\\ Rua do Mat\~ao, 1371, 05508-090  S\~ao Paulo, SP, Brazil}




\email{rivelles@fma.if.usp.br}

\begin{abstract}  
We discuss in a systematic way the gauge theory for a continuous spin particle proposed by Schuster and Toro. We show that it is naturally formulated in a cotangent bundle over Minkowski spacetime where the gauge field depends on the spacetime coordinate ${x^\mu}$ and on a covector $\eta_\mu$.  We discuss how fields can be expanded in $\eta_\mu$ in different ways and how these expansions are related to each other. The field equation has a derivative of a Dirac delta function with support on the $\eta$-hyperboloid $\eta^2+1=0$ and we show how it restricts the dynamics of the gauge field to the $\eta$-hyperboloid and its first neighbourhood. We then show that on-shell the field carries one single irreducible unitary representation of the Poincaré group for a continuous spin particle. We also show how the field can be used to build a set of covariant equations found by Wigner describing the wave function of one-particle states for a continuous spin particle. Finally we show that it is not possible to couple minimally a continuous spin particle to a background abelian gauge field, and make some comments about the coupling to gravity. 
\end{abstract}


\maketitle

\section{Introduction \label{s1}}

It is remarkable that quantum mechanics and Poincaré invariance alone are enough to determine what sort of particles may and may not exist in flat spacetime. The elementary particles are associated with the irreducible unitary representations of the Poincaré group which were classified by Wigner \cite{Wigner:1939cj}. They are characterised by the eigenvalues of $P^2$ and $W^2$, where $W^\mu = \epsilon^{\mu\nu\rho\sigma} P_\nu J_{\rho\sigma}$ is the Pauli-Lubanski vector. In the massive case the states can have integer or half-integer spins. In the massless case we have two classes of representations. Those with $W^2=0$ are the usual helicity states and those with $W^2 = - \rho^2 \not= 0$ give rise to the continuous spin particle (CSP) states with continuous spin $\rho$. In this case, going to a light-cone frame with momentum $k_+\not = 0, k_- = k^i =0$ (with a metric which is mostly minus) the Pauli-Lubanski vector has components $W^+ = 0, \, W^- = - k_+ \epsilon^{ij}J_{ij}$ and $W^i = - k_+ \epsilon^{ij} J_ {j-}$, so that $W^2 = - W^i W^i$ and the helicity operator is $h = - W^-/k_+$. A basis with vectors $|\rho,h>$ which are simultaneously eigenvectors of $W^2$ and $h$, with eigenvalues $-\rho^2$ and $h$, respectively, must satisfy 
\begin{eqnarray}
	W^2 |\rho,h> &=& - \rho^2 |\rho,h>, \qquad \rho^2 > 0, \label{b1} \\
	h |\rho,h>  &=& h |\rho,h>, \qquad \quad \,\, h= 0, \pm 1, \pm 2, \dots  \label{b2} \\
	W_{\pm} |\rho,h> &=& \pm i \rho |\rho,h \pm 1>, \label{b3}
\end{eqnarray}
where $W_\pm = W^1 \pm i W^2$ increases/decreases the helicity by one unit so that the irreducible representation comprises all basis vectors $\{|\rho,h>, h= 0, \pm 1, \pm 2, \dots\}$  and hence it is infinite dimensional. When we take the limit $\rho \rightarrow 0$ we get an infinite number of helicity states with all values of the helicity. This is to be contrasted with the situation where we look for representations with $\rho=0$ which gives rise to the familiar helicity states with a fixed value of $h$. This means that a field theory for a CSP with a smooth limit when  $\rho \rightarrow 0$ should reduce to a massless higher spin (HS) field theory with all values of $h$ being present once. 

While massive and massless particles with $W^2=0$ are found in Nature and can be described by quantum field theories, the same is not true for CSPs. They do not seem to exist and the many attempts to describe them using quantum field theory techniques have failed \cite{Yngvason:1970fy,Iverson:1971hq,Chakrabarti:1971rz,Abbott:1976bb,Hirata:1977ss}. They have been treated using the Bargmann-Wigner equations \cite{Bargmann:1948ck,Hirata:1977ss,Bengtsson:2013vra}, or by proposing covariant equations \cite{Iverson:1971hq,Abbott:1976bb} or derived from higher-dimensional massive HS equations \cite{Bekaert:2005in}. Extensions to higher dimensions and to the supersymmetric case also have been performed \cite{Brink:2002zx} and its connection with tensionless strings studied \cite{Savvidy:2003fx,Mourad:2006xk}. Without a solid field theoretic formulation it is very hard to analyse CSPs interactions. CSPs may not exist, for instance, because they comprise an infinite number of massless states with all possible values of the helicity and there is a huge body of evidence that HS in flat spacetime do not interact (for a review see \cite{Bekaert:2010hw}). On the other side, they seem to have soft emission amplitudes which tend to ordinary low helicity amplitudes at energies larger than $\rho$ \cite{Schuster:2013pxj,Schuster:2013vpr}. The fact that no field theory was known for CSPs is clearly a huge drawback to understand its properties. However, a great advance was achieved recently when an unconventional gauge field theory was proposed by Schuster and Toro \cite{Schuster:2014hca}. It makes use of a gauge field $\Psi(\eta,x)$ depending on the spacetime coordinate $x^\mu$ and an extra coordinate $\eta^\mu$ with an action functional containing Dirac delta functions of $\eta^\mu$. These new features make it hard to understand several aspects of the theory even at the free level case. 

In this paper we want to clarify some important points of the Schuster and Toro proposal. First of all it seems natural to expect that the gauge field depends on an extra coordinate $\eta^\mu$. The one-particle wave function for a CSP derived by Wigner \cite {Wigner:1948zp} depends on the CSP momentum and on an extra variable which is also a 4-vector. Many HS theories formulations make use of a field $\Psi(\eta,x)$ either as a way to manipulate the many indices associated to the HS field (see for instance \cite{Ponomarev:2016jqk}) or sometimes associated to constraints in particle models (see for instance \cite{Bengtsson:2013vra}). In Section \ref{i1} we will show that the role of $\eta^\mu$ is to extend Minkowski spacetime to a cotangent bundle over Minkowski spacetime where the gauge field $\Psi(\eta,x)$ lives. We find that the natural symplectic structure of the cotangent bundle does not seem to be relevant for the gauge theory but the cotangent bundle structure seems to be fundamental when considering curved spacetimes \cite{rivelles}. The gauge field $\Psi(\eta,x)$ is assumed to be analytic in $\eta^\mu$ and we show how it can be expanded in several ways in Section \ref{ssg1}. 

In \cite{Rivelles:2014fsa} we showed how the Schuster and Toro field equations are related to previously proposed field equations for  CSP and HS theories and we also showed how it describes the physical degrees of freedom of a CSP. In order to do that we had to make different gauge choices for each situation turning the reading of the paper somewhat cumbersome. In this paper we will make a single gauge choice for each gauge symmetry so that all intermediate steps become much more transparent. As remarked before the field equation has a very peculiar form involving the derivative of a Dirac delta function $\delta^\prime(\eta^2+1)$. In Section \ref{l1} we show how to deal with this sort of field equation and how it localises the dynamics of the gauge field $\Psi(\eta,x)$ on the $\eta$-hyperboloid $\eta^2+1=0$ and its first neighbourhood.  This will allow us to show in Section \ref{ssi1} that $W^2 \Psi = - \rho^2 \Psi$ up to gauge transformations, without any gauge fixing, generalizing somewhat the results of \cite{Rivelles:2014fsa}. In Section VI we use a new gauge choice to find the physical degrees of freedom improving the derivation presented in \cite{Rivelles:2014fsa}. This result is then used in Section VII to show explicitly the helicity mixing (\ref{b3}) that the CSP states have to satisfy. Moreover, in Section \ref{d16} we show that these degrees of freedom satisfy the Wigner conditions for a CSP, a set of covariant equations for the wave function of one particle states, confirming in an alternative way that we are describing a single CSP. Finally, in the last Section we show that there is no minimal coupling between a CSP and an abelian gauge field and make some comments on the coupling to gravity.

\section{Cotangent Bundle Formulation \label{i1}}

The gauge theory for a CSP can be formulated in a cotangent bundle over Minkowski spacetime. Any field $\Psi(\eta,x)$ depends on the spacetime coordinate $x^\mu$ and on the covector $\eta_\mu$  and it is assumed to be a formal power series in $\eta_\mu$ (the index of $\eta_\mu$ can be raised with the Minkowski metric)
\begin{equation}\label{g2}
	\Psi(\eta,x) = \sum_{n=0}^\infty \frac{1}{n!} \eta^{\mu_1} \dots \eta^{\mu_n} \Psi_{\mu_1 \dots \mu_n}(x),
\end{equation}
where $\Psi_{\mu_1 \dots \mu_n}(x)$ are completely symmetric tensor fields in spacetime which, of course, are expected to be related to the CSP one-particle states. In order to be able to build an action in the cotangent bundle it is an integration procedure is needed which allow us to get meaningful actions for the components of the $\Psi(\eta,x)$. Such a procedure was developed in \cite{Segal:2001qq} and requires the use of distributions localised on the $\eta$-hyperboloid $\eta^2+1=0$ and its neighbourhoods. Consider the integral $\int d^4\eta \, \theta(\eta^2+1)$, where $\theta(\alpha)$ is the usual step function vanishing for $\alpha<0$. After a Wick rotation the integral is well defined and is proportional to the volume of the sphere with unit radius. Integrals of the form $\int d^4\eta \, \theta(\eta^2+1) \varphi[\Psi(\eta,x)]$ are also well defined and, taking into account the expansion (\ref{g2}), they will be proportional to a sum of contracted $\Psi(\eta,x)$ components. This can be generalised by considering derivatives of the $\theta$ function so that $\int d^4\eta \, \delta^{(n)}(\eta^2+1) \varphi[\Psi(\eta,x)]$, where $\delta^{(n)}$ is the $n$-th derivative of the delta function with respect to its argument, is also well defined. Besides that, these integral expressions allow us to perform integration by parts in $\eta^\mu$ so that self-adjoint operators can be defined in the standard way. The Wick rotation is only needed if we want to compute the integrals over $\eta^\mu$ in terms of the component fields as done in \cite{Rivelles:2014fsa} to get the Fronsdal action when $\rho=0$. In all other situations the manipulations are independent of the Wick rotation. Is is also important to note that there is nothing special about the $\eta$-hyperboloid $\eta^2+1=0$. We could have started with the hyperboloid $\eta^2+\mu^2=0$ and absorbed $\mu$ through a redefinition of $\eta^\mu$ and $\Psi(\eta,x)$. 

The action for a CSP is given by \cite{Schuster:2014hca}
\begin{equation}\label{2.1}
	S = \frac{1}{2} \int d^4 x \,\, d^4\eta \left[ \delta^\prime(\eta^2+1) (\partial_x \Psi(\eta,x))^2 + \frac{1}{2} \delta(\eta^2+1) \left( \Delta \Psi(\eta,x) \right)^2 \right],
\end{equation}
where $\Delta = \partial_\eta \cdot \partial_x + \rho$ and $\delta^\prime$ is the derivative of the delta function with respect to its argument. The presence of delta functions gives rise to unfamiliar field equations which are localised on the hyperboloid $\eta^2+1=0$ and its first neighbourhood as we will discuss in detail in the next sections. The action is invariant under Lorentz transformations since $\eta_\mu$ is a covector and its generator is $J_{\mu\nu} = i x_{[\mu} \partial_{x\nu]} + i \eta_{[\mu} \partial_{\eta\nu]}$. It is also invariant under translations in spacetime, with generators $P_{\mu} = i \partial_{x\mu}$, but not translations in $\eta^\mu$. Besides, it is invariant, up to spacetime surface terms, under the global transformation
\begin{equation}
	\delta \Psi = - \omega^{\mu\nu} \eta_\mu \partial_{x\nu} \Psi, \label{i2}
\end{equation}
with the constant parameter $\omega^{\mu\nu}$ being antisymmetric. This symmetry of the action is a consequence of the transformation 
\begin{eqnarray}
	\delta x^\mu &=& \omega^{\mu\nu} \eta_\nu, \label{r1} \\
	\delta \eta^\mu &=& 0, \label{r2}
\end{eqnarray}
which is an $\eta^\mu$ dependent translation along $x^\mu$ for fixed $\eta^\mu$ with generator  $N_{\mu\nu} = i \eta_{[\mu} \partial_{x\nu]}$. The set of generators $(P_{\mu}, J_{\mu\nu}, N_{\mu\nu})$ form a closed algebra where $N_{\mu\nu}$ commutes with $P_{\mu}$ and with itself while transforming as an antisymmetric tensor under Lorentz transformations. Notice that 
 $(P_{\mu}, J_{\mu\nu}, N_{\mu\nu})$ gives rise to conserved Noether charges, while the irreducible unitary representations for a CSP are characterised by the Casimir operators $P^2$ and $W^2$ of the Poincaré group, and not of the enlarged group with generators $(P_{\mu}, J_{\mu\nu}, N_{\mu\nu})$. Also, this new global symmetry 
does not preserve the natural symplectic structure of the cotangent bundle $\Omega = d\eta_\mu \wedge dx^\mu$ so that $x^\mu$ and $\eta_\mu$ are not canonical variables. However, it preserves the infinitesimal cotangent bundle volume in (\ref{2.1}) due to the antisymmetry of $\omega^{\mu\nu}$. Thus, the symplectic structure seems to play no role in this case\footnote{HS theories on cotangent bundles have been proposed in \cite{Fronsdal:1979gk} and its symplectic structure has been exploited in several situations. See, for instance, \cite{Ponomarev:2013mqa}  and references therein.} but the cotangent bundle framework is useful when CSPs in curved spacetime are considered \cite{rivelles}.  

The action is also invariant under the local transformations
\begin{equation}\label{2.2}
	\delta \Psi(\eta,x) = \left( \eta\cdot\partial_x - \frac{1}{2} (\eta^2+1) \Delta \right) \epsilon(\eta,x)  + \frac{1}{4} (\eta^2+1)^2 \chi(\eta,x),
\end{equation}
where $\epsilon(\eta,x)$ and $\chi(\eta,x)$ are the local parameters. 
As we shall see the $\eta^\mu$ expansion of $\Psi(\eta,x)$ provides a highly redundant description in terms of spacetime fields and the $\chi$ symmetry can be used to simplify the expansion of $\Psi$. The transformation with parameter $\epsilon(\eta,x)$ is a gauge transformation and it reduces to the usual Fronsdal gauge transformations for massless HS fields when $\rho=0$. The symmetries (\ref{2.2}) are reducible \cite{Rivelles:2014fsa} since 
\begin{eqnarray}\label{2.3}
	\delta \epsilon &=& \frac{1}{2} (\eta^2+1) \Lambda(\eta,x), \\ \label{2.4}
	\delta \chi &=& \Delta \Lambda(\eta,x),
\end{eqnarray}
leave the RHS of (\ref{2.2}) invariant. This symmetry can be used to simplify the $\eta^\mu$ expansion of $\epsilon$ as we will see shortly. 

\section{Expansion of \texorpdfstring{$\Psi(\eta,x)$}{}   \label{ssg1}} 

We assumed that $\Psi(\eta,x)$ is analytic in $\eta^\mu$ and as such it can be expanded in several ways. The formal power series (\ref{g2}) presents some interesting properties. We can decompose each $\Psi_{\mu_1 \dots \mu_n}(x)$ in its trace and traceless parts and each trace will contribute with a factor $\eta^2$ in the sum (\ref{g2}). Each trace can now be decomposed in an analogous way generating an extra factor of $\eta^2$ and so on. All $(\eta^2)^n$ factors can then be grouped together in such a way that $\Psi(\eta,x)$ can be written as
\begin{eqnarray}\label{g3}
	\Psi(\eta,x) &=& \sum_{n=0}^\infty \frac{1}{n!} (\eta^2)^n \varphi^T_n(\eta,x), \\
	\varphi^T_n(\eta,x)&=& \sum_{p=0}^\infty \frac{1}{p!} \eta^{\mu_1} \dots \eta^{\mu_p} \varphi^T_{n,\mu_1\dots\mu_p}(x), \label{g4}
\end{eqnarray}
where $\varphi^T_{n,\mu_1\dots\mu_p}(x)$ is traceless. We can now write $\eta^2 = (\eta^2 +1) - 1$ and use the binomial expansion to get 
\begin{eqnarray}\label{g5}
	\Psi(\eta,x) &=& \sum_{n=0}^\infty \frac{1}{n!} (\eta^2 + 1)^n \phi^T_n(\eta,x), \\
	\phi^T_n(\eta,x)&=& \sum_{p=0}^\infty \frac{(-1)^p}{2^{2(p+n)} p!} \Phi^T_{p+n}(\eta,x), \label{g6} \\
	\Phi^T_{p}(\eta,x) &=& \sum_{s=0}^\infty \frac{s+1}{(s+p+1)!} \eta^{\mu_1} \dots \eta^{\mu_s} \Psi^{\overbrace{\prime \dots \prime}^{p \,\, \text{times}}T}_{\mu_1 \dots \mu_s}(x), \label{k1} 
\end{eqnarray}
where $\Psi^{\overbrace{\prime \dots \prime}^{p \,\, \text{times}}T}_{\mu_1 \dots \mu_s}(x)$ is the traceless part of the $p$-th trace of $\Psi_{\mu_1 \dots \mu_{s+2p}} (x)$. This procedure was used in \cite{Segal:2001qq} to formulate a higher spin theory for all integer spins in terms of two traceless fields of the form (\ref{g2}). 


In \cite{Rivelles:2014fsa} we wanted to make contact with other CSP and HS formulations. To do that and to use the reducibility in (\ref{2.3}) and (\ref{2.4}) it was found that it is better to expand $\Psi(\eta,x)$ as 
\begin{eqnarray}\label{2.5}
	\Psi(\eta,x) &=& \sum_{n=0}^{\infty} \frac{1}{n!} (\eta^2+1)^n \psi_n(\eta,x), \\
\label{2.6}
	\psi_n(\eta,x) &=& \sum_{s=0}^\infty \frac{1}{s!} \eta^{\mu_1}\dots\eta^{\mu_s} \psi_{n,\mu_1\cdots\mu_s}(x),
\end{eqnarray}
where $\psi_{n,\mu_1\cdots\mu_s}(x)$ is completely symmetric and does not satisfy any traceless condition whatsoever. This decomposition, however, is not unique \cite{Rivelles:2014fsa}. If we transform $\psi_n(\eta,x)$ as 
\begin{equation}
	\delta \psi_n(\eta,x) = \sum_{p=1}^\infty \frac{n!}{(n+p)!} (\eta^2+1)^p \,\, \Xi_{n,n+p}(\eta,x) - \sum_{p=0}^{n-1} \Xi_{n,p}(\eta,x), \label{2.7}
\end{equation}
with $\Xi_{n,p}(\eta,x)$ symmetric in $n$ and $p$ then $\Psi(\eta,x)$ is invariant. If we use the expansion (\ref{2.5}) in the action ({\ref{2.1}) the transformations (\ref{2.7}) will trivially leave the action invariant. The transformations (\ref{2.7}) do not constitute a gauge symmetry since they are not removing any degree of freedom. Its role is just to reshuffle traces among the $\psi_n(\eta,x)$ and, as shown in \cite{Rivelles:2014fsa}, it can be used to extend the field equation off the $\eta$-hyperboloid. 

To analyse the consequences of the action (\ref{2.1}) we can consider $\Psi(\eta,x)$ expanded as in (\ref{g5}), or as in (\ref{2.5}) or even with no expansion whatsoever, depending on our aim. Taking into account the $\chi$ symmetry in (\ref{2.2}) it is natural to expand $\Psi(\eta,x)$ as in (\ref{2.5}) since it allow us to gauge away all terms in the expansion of $\Psi(\eta,x)$ but the first two. If instead we use the expansion (\ref{g5}) the components are traceless but they are hard to deal with because they must satisfy the constraint $\Box_\eta \psi^T_n(\eta,x)=0$.  Then it seems natural to start with the expansion (\ref{2.5}) since no constraints are required. 

The introduction of an extra coordinate is quite useful when regarding HS theories. In terms of an expansion like (\ref{2.5}) the Fronsdal equation for integer spin $s$ \cite{Fronsdal:1978rb} is 
\begin{equation}
	\left( \Box_x - \eta\cdot\partial_x \partial_\eta\cdot\partial_x + \frac{1}{2} (\eta\cdot\partial_x)^2 \Box_\eta \right) \psi(\eta,x) = 0, \label{h1}
\end{equation}
with the condition $(\eta\cdot\partial_\eta - s) \psi=0$ to select just the field with spin $s$, while the double traceless condition is written  as $\Box^2_\eta \psi=0$ and the gauge transformation takes the form $\delta\psi = \eta\cdot\partial_x \epsilon$. This equation was  considered before but no action from which they could be derived was known. As shown in \cite{Rivelles:2014fsa} the action is given by (\ref{2.1}) with $\rho=0$

\section{The Delta Function Structure of the Field Equation \label{l1}}

The field equation obtained from (\ref{2.1})
\begin{equation}
	\delta^\prime(\eta^2+1) \left( \Box_x \Psi - \eta\cdot\partial_x \Delta\Psi + \frac{1}{2} (\eta^2+1) \Delta^2\Psi \right) = 0, \label{i3}
\end{equation}
has an unusual structure involving the derivative of a Dirac delta function. It implies that the derivative of the term between parentheses with respect to $\eta^2+1$ evaluated at $\eta^2+1=0$ vanishes. This means that we have a differential equation for $\Psi(\eta,x)$ involving a $\eta^\mu$ coordinate which is constrained by $\eta^2+1=0$. This is a nuisance since calculations soon become too complicated. For instance, $\partial \eta^\mu/\partial \eta^\nu$ is now a projector and no longer a simple $\delta^\mu_\nu$. So we have 
to try to handle the delta function structure in another way in order to have a conventional field equation,  involving just $\Psi(\eta,x)$ and its derivatives without any constraint on $\eta^\mu$. 

To start let us consider a simpler equation in just one dimension $\delta(x) f(x) = 0$. Since the delta function has support at $x=0$ the solution is any function $f(x)$ which vanishes at $x=0$. If $f(x)$ is not required to satisfy any other condition then $f(x)$ is not unique and we can pick up any $f(x)$ vanishing at $x=0$, in particular $f(x)=0$. An alternative way to take into account the fact that the delta function has support at $x=0$ is to notice that $\delta(x) f(x) = 0$ is invariant under the local transformation $f(x) \rightarrow f(x) + x \, \Lambda(x)$. This resembles a gauge transformation but it is just a consequence of the delta function structure of the equation. We can now solve the equation as $f(x) = x \, \omega(x)$ with $\omega(x)$ finite at $x=0$. In the context of gauge theory we would call this solution as pure gauge since  we can always find a $\Lambda(x)$ which leads to the solution $f(x)=0$. Of course, our local transformation is not a gauge transformation since we are not dealing with redundant degrees of freedom in a gauge theory. Here, the local transformation  means that $f(x)$ is restricted to the support of the delta function and may be extended outside the support as we like.


The field equation (\ref{i3}) has a similar delta function structure since it can be written as 
\begin{eqnarray}
	&&\delta^\prime(\eta^2+1) A(\eta,x) = 0, \label{q2} \\
	&&A(\eta,x) = \Box_x \Psi - \eta\cdot\partial_x \Delta\Psi + \frac{1}{2} (\eta^2+1) \Delta^2\Psi. \label{q3}
\end{eqnarray}
This means that $A^\prime(\eta,x)$ (the first derivative of $A(\eta,x)$ with respect to $\eta^2+1$) vanishes on the $\eta$-hyperboloid. We can then extend the solution beyond the first neighbourhood of the $\eta$-hyperboloid as $A(\eta,x)=0$ providing a differential equation for $\Psi(\eta,x)$. This means that the dynamics of $\Psi(\eta,x)$ takes place only on the $\eta$-hyperboloid and its first neighbourhood since (\ref{i3}) is not strong enough to fix the dynamics beyond the first neighbourhood of the $\eta$-hyperboloid. In conclusion the field equation (\ref{i3}) describes the dynamics of $\Psi(\eta,x)$ not on all of the cotangent bundle but only on the $\eta$-hyperboloid and its first neighbourhood. 

Alternatively we can consider the solution of (\ref{q2}) $A(\eta,x) = (\eta^2+1)^2 \omega(\eta,x)$ with $\omega(\eta,x)$ finite on the $\eta$-hyperboloid and its first neighbourhood. 
As in the one dimensional case it is invariant under the local transformation 
\begin{equation}
	 A(\eta,x) \rightarrow A(\eta,x) + (\eta^2+1)^2 \theta(\eta,x) \label{q1}
\end{equation}
where $\theta(\eta,x)$ is any function of $\Psi(\eta,x)$ and its derivatives. Again, this shows that $A(\eta,x)$ is invariant on the $\eta$-hyperboloid and its first neighbourhood and arbitrary outside it. But as we saw earlier the action (\ref{2.1}) and consequently the field equation (\ref{i3}) are invariant under the local transformations (\ref{2.2}) with parameters $\epsilon(\eta,x)$ and $\chi(\eta,x)$,  while the combination $A(\eta,x)$ is not invariant
\begin{eqnarray} 
	\delta_\epsilon A(\eta,x) &=& - \frac{1}{4} (\eta^2+1)^2 \Delta^3 \epsilon, \label{i7} \\
	\delta_\chi A(\eta,x) &=& \frac{3}{4} (\eta^2+1)^2  ( \Box_x \chi + \eta\cdot\partial_x \Delta\chi + \frac{1}{6} (\eta^2+1) \Delta^2 \chi ). \label{i5}
\end{eqnarray}
These transformations, of course, have the same $\eta^2+1$ structure of the solution $A(\eta,x) = (\eta^2+1)^2 \omega(\eta,x)$ so that $A(\eta,x)$ is invariant under (\ref{2.2}) on the $\eta$-hyperboloid and its first neighbourhood up to a $\theta$ transformation (\ref{q1}). 
Then we can still can use the $\epsilon(\eta,x)$ transformation to deal with the gauge transformation for the components of $\Psi(\eta,x)$, the $\chi(\eta,x)$ transformation to simplify the $\eta^2+1$ expansion of $\Psi(\eta,x)$, and the $\theta(\eta,x)$ transformation to choose $A(\eta,x)=0$ all over of the cotangent space.


In this way we can solve the delta function constraint of (\ref{i3}) as 
\begin{equation}
	\Box_x \Psi - \eta\cdot\partial_x \Delta\Psi + \frac{1}{2} (\eta^2+1) \Delta^2\Psi = 0, \label{i6}
\end{equation}
everywhere in $\eta$-space but having in mind that only solutions on the $\eta$-hyperboloid and its first neighbourhood  have to be taken into account. Hence this allows us to perform  calculations on the cotangent bundle with $\eta^\mu$ unconstrained. 

\section{The Casimir Operator \texorpdfstring{$W^2$}{} \label{ssi1}}

To find the irreducible representation carried by $\Psi(\eta,x)$ we have to evaluate the eigenvalue of the square of the Pauli-Lubanski operator on-shell. 
To compute it we must realise the Poincaré generators on the cotangent bundle. $P_\mu$ is realised as usual as a spacetime derivative. However, $J_{\mu\nu}$, as we saw earlier, has a new term $i \eta_{[\mu} \partial_{\eta \nu]}$.  
Then the Pauli-Lubanski vector acts on $\Psi(\eta,x)$ as $ W^\mu \Psi = - \epsilon^{\mu\nu\rho\sigma} \partial_{x\nu} \eta_\rho \partial_{\eta \sigma} \Psi$ so that 
\begin{equation}
	W^2 \Psi = \left[\eta\cdot\partial_\eta(1+\eta\cdot\partial_\eta) \Box_x  -  \eta^2 \Box_\eta \Box_x  - 2\eta\cdot\partial_\eta \, \eta\cdot\partial_x \, \partial_\eta\cdot\partial_x  
	 + (\eta\cdot \partial_x)^2 \Box_\eta  + \eta^2 (\partial_\eta \cdot \partial_x)^2\right] \Psi. \label{ii2}
\end{equation}
Using the field equation (\ref{i6}) we find that
\begin{equation}
	W^2 \Psi = - \rho^2 \Psi + \delta_\epsilon \Psi + \delta_\chi \Psi, \label{i8}
\end{equation}
with
\begin{eqnarray}
	\epsilon &=&  \eta\cdot\partial_\eta (1+\eta\cdot\partial_\eta) \Delta\Psi + 2\rho (1+\eta\cdot\partial_\eta)\Psi + (\eta\cdot\partial_x  + \Delta) \Box_\eta\Psi, \label{i9} \\
	\chi &=&  \Box_\eta \Delta^2 \Psi. \label{i10}
\end{eqnarray}
Then, up to local transformations, $\Psi$ carries an irreducible representation of the Poincaré group with $W^2= -\rho^2$ as expected for a CSP. As explained in the previous section this is true only on the $\eta$-hyperboloid and its first neighbourhood and not on all of the cotangent bundle. This same result, the computation of $W^2$ without gauge fixing $\epsilon$ and $\chi$, was obtained in \cite{Rivelles:2014fsa} explicitly on the $\eta$-hyperboloid. 

\section{Physical Degrees of Freedom}\label{c1}

To unravel the physical degrees of freedom we start with $\Psi(\eta,x)$ in the form (\ref{2.5}) and use the $\chi$ symmetry in  (\ref{2.2}) to gauge away all $\psi_n(\eta,x)$ with $n \ge 2$ \cite{Rivelles:2014fsa} 
\begin{equation}\label{c2}
	\Psi(\eta,x) = \psi_0(\eta,x) + (\eta^2+1) \psi_1(\eta,x).
\end{equation}
We can now use the same expansion (\ref{2.5}) for the gauge parameter $\epsilon(\eta,x)$ and use the $\Lambda$ symmetry in (\ref{2.3}) to gauge away all $\epsilon_n(\eta,x)$ with $n \ge 1$ so that $\epsilon(\eta,x) = \epsilon_0(\eta,x)$ \cite{Rivelles:2014fsa}. At this point there is no traceless condition on $\psi_0, \psi_1$ and $\epsilon_0$ so that we still have the $\Xi$ symmetry (\ref{2.7}) which, together with the $\epsilon$ gauge transformation in (\ref{2.2}), gives  
\begin{eqnarray}
	\delta \psi_0 &=& \eta\cdot\partial_x \epsilon_0 + (\eta^2+1) \Xi, \label{c3} \\
	\delta \psi_1 &=& - \frac{1}{2} \Delta \epsilon_0 - \Xi. \label{c4}
\end{eqnarray}

The field equation (\ref{i3}) can be rewritten as 
\begin{equation}
	\delta^\prime(\eta^2+1) \left[ A(\eta,x) + 2 (\eta^2+1) B(\eta,x) \right] = 0, \label{c5}
\end{equation}
where now
\begin{eqnarray}
 A(\eta,x) &=&  \Box_x \psi_0 - \eta \cdot \partial_x \Delta \psi_0 - 2 (\eta \cdot \partial_x)^2 \psi_1, \label{c.6} \\
 B(\eta,x) &=& \Box_x \psi_1 + \frac{1}{2} \eta\cdot\partial_x \Delta \psi_1 + \frac{1}{4} \Delta^2 \psi_0.  \label{c.7}
\end{eqnarray}
Notice that $A$ and $B$ are not independent since they are related by
\begin{equation}
	\Delta A = - 4\eta\cdot\partial_x B. \label{2.18a}
\end{equation}
Also, they are invariant under an $\epsilon$ transformation but not under a $\Xi$ transformation
\begin{eqnarray}
	&&\delta_\Xi A(\eta,x) = (\eta^2+1) (\Box_x - \eta\cdot\partial_x \Delta) \Xi, \label{w1} \\
	&&\delta_\Xi B(\eta,x) = -\frac{1}{2} (\Box_x - \eta\cdot\partial_x \Delta) \Xi + \frac{1}{4} \Delta^2 \Xi. \label{w2}
\end{eqnarray}

As discussed in Section \ref{l1} the general solution of (\ref{c5}) is $A(\eta,x) + 2 (\eta^2+1) B(\eta,x) = (\eta^2+1)^2 \omega(\eta,x)$, 
and it can always be chosen to vanish, so that
\begin{equation}
	A(\eta,x) + 2 (\eta^2+1) B(\eta,x)=0, \label{t1}
\end{equation}
everywhere in $\eta$-space. Since $A(\eta,x)$ and $B(\eta,x)$ are not invariant under a $\Xi$ transformation we find from (\ref{w1}) and (\ref{w2}) that there is still a residual $\Xi$ transformation satisfying $(\eta^2+1)\Delta^2\Xi=0$, so that $\Delta^2\Xi=0$. We can still use the residual $\Xi$ transformation to set $A(\eta,x)=0$ implying that the residual $\Xi$ transformation is further constrained by $ (\Box_x - \eta\cdot\partial_x \Delta) \Xi = 0$. On the other side, using (\ref{2.18a}) we find that $\eta\cdot\partial_x B=0$, so that $B(\eta,x)=0$ up to zero modes which are not relevant since we are looking for representations of the Poincar\'e group with light-like momentum. We have then shown that $A(\eta,x)=B(\eta,x)=0$ everywhere on $\eta$-space so that the field equation (\ref{c5}) becomes 
\begin{eqnarray}
  \Box_x \psi_0 - \eta \cdot \partial_x \Delta \psi_0 - 2 (\eta \cdot \partial_x)^2 \psi_1 = 0, \label{m5} \\
  \Box_x \psi_1 + \frac{1}{2} \eta\cdot\partial_x \Delta \psi_1 + \frac{1}{4} \Delta^2 \psi_0 = 0,  \label{m6}
\end{eqnarray}
on all of $\eta$-space. There is still a residual $\Xi$ symmetry in (\ref{c3}) and (\ref{c4}) with the parameter satisfying
\begin{equation}
 (\Box_x - \eta\cdot\partial_x \Delta)\Xi = \Delta^2 \Xi = 0. \label{j1}
\end{equation}


As we have seen, (\ref{t1}) is a solution of (\ref{c5}) which has support in the first neighbourhood of the $\eta$-hyperboloid. We have shown that it is always possible to choose $A(\eta,x)=B(\eta,x)=0$. But this solution is stronger than (\ref{t1}) and in fact has support on the $\eta$-hyperboloid and not on its first neighbourhood. This can be seem by multiplying (\ref{c5}) by $\eta^2+1$ to get $\delta(\eta^2+1) A(\eta,x)=0$ which has support on the $\eta$-hyperboloid. The solution is $A(\eta,x)=(\eta^2+1) \omega_A(\eta,x)$ and we can then choose $A(\eta,x)=0$. Replacing this solution in (\ref{c5}) we find that $\delta(\eta^2+1) B(\eta,x)=0$ which also has support on the $\eta$-hyperboloid. The solution is $B(\eta,x)=(\eta^2+1) \omega_B(\eta,x)$ and we can also choose $B(\eta,x)=0$. A solution in the first neighbourhood should have $A(\eta,x)\not= 0$. 

To analyse the physical degrees of freedom we start with the harmonic gauge choice
\begin{equation}
  \Delta\psi_0 +	2\eta\cdot\partial_x \psi_1 = 0. \label{j2}
\end{equation}
When used in (\ref{m5}) and (\ref{m6}) it implies that $\Box_x \psi_0 = \Box_x \psi_1 = 0$ and when we take an $\epsilon$ gauge transformation of (\ref{j2}) we get $\Box_x \epsilon_0=0$. Considering now the residual $\Xi$ transformation on (\ref{j2}) we find that $\Box_x\Xi = \Delta \Xi =0$. 

We still have room for a further gauge choice since $\epsilon_0$ satisfies only $\Box_x\epsilon_0=0$ so that we choose $\Delta\psi_0$=0. Using (\ref{c3}) this means that there is a residual $\epsilon$ gauge symmetry with $\Delta\epsilon_0=0$. Using now (\ref{j2}) we find that $\psi_1=0$ and from (\ref{c4}) we get $\Xi=0$ so that the $\Xi$ symmetry is completely fixed. In summary we are  left with $\psi_0$ satisfying $\Box_x \psi_0 = \Delta\psi_0 = 0$ and a residual gauge transformation $\delta\psi_0 = \eta\cdot\partial_x \epsilon_0$ with $\Box_x \epsilon_0 = \Delta\epsilon_0 = 0$. 

In components $\Delta\psi_0$=0 can be written in momentum space as
\begin{equation}
	 i k\cdot\tilde{\psi}_{0\mu_1\dots \mu_n}(k) + \rho \tilde{\psi}_{0\mu_1\dots \mu_n}(k) = 0. \label{j3}
\end{equation}
In a Lorentz frame where the light-cone components of the momentum satisfy $k_- = k_i =0$, $(i=1,2)$ and using the notation $\tilde{\psi}_{\underbrace{+\dots +}_{p \,\, \text{times}} \underbrace{-\dots -}_{q \,\, \text{times}} i_1 \dots i_n}(k) = \tilde{\psi}_{ (+)^p (-)^q (i)^n}(k)$ for the light-cone components of $\tilde{\psi}_{\mu_1\dots\mu_{p+q+n}}(k)$, (\ref{j3}) can be rewritten as 
\begin{equation}
	 ik_+ \tilde{\psi}_{0 (+)^p (-)^{q+1} (i)^n} + \rho \tilde{\psi}_{0 (+)^p (-)^q (i)^n} =0, \qquad p,q,n \ge 0. \label{j4}
\end{equation}
This equation can be solved for the $-$ components as
\begin{equation}
	\tilde{\psi}_{0 (+)^p(-)^q (i)^n} = \left( -\frac{\rho}{ik_+} \right)^q \tilde{\psi}_{0 (+)^p (i)^n}, \qquad p,q,n \ge 0,  \label{j5}
\end{equation}
so that the independent components of $\psi_0$ are $\tilde{\psi}_{0 (+)^p (i)^n}$. Since $\epsilon_0$ satisfies the same equation $\Delta\epsilon_0=0$, we also find that its independent components are $\tilde{\epsilon}_{0 (+)^p (i)^n}$.

The residual $\epsilon$ gauge transformation for ${\psi}_0$ can be written for the Fourier components  as 
\begin{equation}
	\delta \tilde{\psi}_{0 \mu_1 \dots \mu_n}(k) = \frac{1}{(n-1)!} ik_{(\mu_1} \tilde{\epsilon}_{\mu_2 \dots \mu_n)}(k), \label{d8}
\end{equation}
which can then be cast into the form
\begin{equation}
	\delta \tilde{\psi}_{0 (+)^p (-)^q (i)^n } = p ik_+ \tilde{\epsilon}_{ (+)^{p-1} (-)^q (i)^n}, \qquad p,q,n \ge 0. \label{j7}
\end{equation}
For $p=0$ we find that $\tilde{\psi}_{0 (-)^q (i)^n}, \,\, q,n \ge 0$, are gauge invariant and because of (\ref{j5}) the independent components $\tilde{\psi}_{0 (i)^n}$ are also gauge invariant. For $p \ge 1$ it is possible to gauge away all $\tilde{\psi}_{0 (+)^p (-)^q (i)^n}, \,\, p\ge 1, q,n \ge 0$ so that all components of $\tilde{\epsilon}$ are used and the gauge is completely fixed. Summarizing, all $+$ components of $\psi_0$ can be gauged away, all $-$  components can be expressed in terms of the $i$ components through (\ref{j5}) and all $i$ components are gauge invariant. We have then found that the physical degrees of freedom are described by $\tilde{\psi}_{0 i_1 \dots i_n}(k)$. 

Up to now we have been handling the equations in the cotangent bundle. We have found that the physical degrees of freedom are carried by the components $\tilde{\psi}_{0 i_1 \dots i_n}(k)$ of $\tilde{\psi}_0(\eta,x)$ and they describe all integer helicities each one appearing an infinite number of times since all traces of $\tilde{\psi}_{0 i_1 \dots i_n}(k)$ are present. Taking into account that (\ref{m5}) and (\ref{m6}) hold on the $\eta$-hyperboloid 
we have now to restrict our solution to it. Since $\psi_1=0$ then (\ref{c2}) reduces to  
\begin{equation}
	\Psi(\eta,x) = \psi_0(\eta,x) = \sum_{n=0}^\infty \frac{1}{n!} \eta^\mu_1 \dots \eta^\mu_n \psi_{0 \mu_1 \dots \mu_n}, \label{j8}
\end{equation}
on the $\eta$-hyperboloid with all $+$ components vanishing and the $-$ components given by (\ref{j5}). Since (\ref{j8}) has the form (\ref{g2}) it can be rewritten as (\ref{g5}) and since we are on the $\eta$-hyperboloid all terms in the sum (\ref{g5}) vanish except for the first one so that on the hyperboloid ${\psi}_0(\eta,x) = \phi^T_0(\eta,x)$.  
Since the components of $\phi^T_0(\eta,x)$ are traceless we are left with an infinite set of traceless spacetime fields so that the physical degrees of freedom on the hyperboloid have all integer helicities but now each helicity appears just once, as expected for a single CSP. The traceless components of $\phi^T_0(\eta,x)$, on its turn, can be computed in terms of the components of $\psi_0(\eta,x)$ using (\ref{g6}) and (\ref{k1}). 

Let us now consider the limit $\rho\rightarrow 0$. The field equations (\ref{m5}) and (\ref{m6}) and the gauge condition (\ref{j2}) are regular in the limit. The solution of the gauge condition is also regular and (\ref{j5}) shows that all $-$ components of $\psi_0$ vanish as expected for Fronsdal fields. The gauge transformation (\ref{d8}) does not depend on $\rho$ and we get same results as for $\rho\not= 0$. The physical degrees of freedom are still described by $\tilde{\psi}_{0 i_1 \dots i_n}$. Going to the $\eta$-hyperboloid does not involve $\rho$ and we get the traceless condition in the same way as for $\rho\not= 0$. Therefore, when $\rho\rightarrow 0$, we find an infinite tower of Fronsdal massless fields for all integer spins living on the $\eta$-hyperboloid, as expected.

\section{\texorpdfstring{$W_\pm$}{} and helicity mixing \label{k2}}

In order to show (\ref{b3}) we must first find combinations of the $\tilde{\psi}_0(\eta,x)$ components which have well defined helicity. To do that we must introduce some helicity notation which, unfortunately, makes use of same symbols used in the light-cone notation and this may  cause some confusion. We apologise for that in advance.  

Let us consider the vector component of $\tilde{\psi}_0(\eta,k)$, that is $\tilde{\psi}_{0\mu}(k)$. As we saw in the previous Section, $\tilde{\psi}_{0+}(k)=0$ and $\tilde{\psi}_{0-}(k)$ is to be expressed in terms of $\tilde{\psi}_{0i}(k)$ through (\ref{j5}) so that the independent components are $\tilde{\psi}_{0 i}(k), \,\, i=1,2$. From now on we will no longer use the light-cone components $\tilde{\psi}_{0\pm}(k)$ and we will introduce the helicity notation for $\tilde{\psi}_{0i}(k)$ as 
\begin{equation}
	\tilde{\psi}_\pm = - \tilde{\psi}^\pm \equiv \frac{1}{\sqrt{2}} (\tilde{\psi}_{0 1} \pm i\tilde{\psi}_{0 2} ).  \label{k5}
\end{equation}
Recalling that $\tilde{\psi}_{0 \mu}(k)$ is the Fourier transformed of $\psi_{0\mu}(x)$ we also have $\tilde{\psi}^\dagger_\pm(k) = \tilde{\psi}_\mp(-k)$. We stress that $\tilde{\psi}_\pm(k)$ are no longer the light-cone components of $\tilde{\psi}_{0\mu}(k)$ but its $\pm$ helicity components. We can do the same decomposition for a completely symmetric tensor $\tilde{\psi}_{0 \mu_1\dots\mu_s}(k)$ which is denoted $\tilde{\psi}^{(s_+,s_-)}(k)$, where $s_\pm$ is number of $\pm$ indices that it carries and $s_+ + s_- =s$ is the tensor order. If $s_+=0$ or $s_-=0$ the tensor is traceless. 

For $\eta^\mu$, with light-cone components $(\eta^+,\eta^-,\eta^i)$, we define
\begin{equation}
	\eta_\pm \equiv \frac{1}{\sqrt{2}} ( \eta_1 \pm i\eta_2 ), \label{k6}
\end{equation}
so that $\eta^\dagger_\pm = \eta_\mp$ and we never use downstairs light-cone components for $\eta^\mu$. Derivatives with respect to $\eta_\pm$ are denoted by $\partial/\partial\eta_\pm$. With this notation we can write the gauge fixed $\tilde{\psi}_0(\eta,k)$ as
\begin{equation}
	\tilde{\psi}_0 (\eta,k) = e^{\frac{i\rho}{k_+}\eta^-} \sum_{s_\pm=0}^{\infty} \frac{1}{s_+! \, s_-!} \eta_+^{s_-} \eta_-^{s_+} \tilde{\psi}^{(s_+,s_-)} (k), \label{n3}
\end{equation}
where (\ref{j5}) was used to eliminate the $-$ components of $\tilde{\psi}_0(\eta,k)$. Notice that its $\eta^\mu$ dependence is only through $\eta^-$, $\eta_+$ and $\eta_-$ since all dependence on the light-cone component $\eta^+$ was gauged away. 

The helicity operator is $h = - W^-/k_+$, where $W^-$ is a light-cone component of $W^\mu$, and it can be written as $ h=h_{(\eta)} + {h}_{(s)}$, where  $h_{(\eta)}$ acts on $\eta^\mu$ as 
\begin{equation}
	h_{(\eta)} = \eta_+ \frac{\partial}{\partial\eta_+} - \eta_- \frac{\partial}{\partial\eta_-}, \label{k7}
\end{equation}
while the spin part ${h}_{(s)}$ acts on $\tilde{\psi}^{(s_+,s_-)}(k)$ as 
\begin{equation}
	{h}_{(s)} \tilde{\psi}^{(s_+,s_-)}(k) = (s_+ - s_-) \tilde{\psi}^{(s_+,s_-)}(k). \label{k8}
\end{equation}
Then the helicity of $\tilde{\psi}^{(s_+,s_-)}(k)$ is $h = s_+ - s_-$ while from (\ref{n3}) we find that $h \tilde{\psi}_0(\eta,k) = 0$ as expected. This happens because ${\psi}_0(\eta,x)$ carries no overall spacetime index, since all of them are contracted as can be seen in (\ref{j8}). 

We can now write the components $W^1$ and $W^2$ of the Pauli-Lubanski vector as $W_\pm = W^1 \pm i W^2$ and, as for the helicity $h$, we split it as $W_\pm = W_{(\eta)\pm} + {W}_{(s)\pm}$ where $W_{(\eta)\pm}$ acts on $\eta^\mu$ as 
\begin{equation}
	W_{(\eta)\pm} = \mp\frac{k_+}{\sqrt{2}} \left( \eta_\pm \frac{\partial}{\partial\eta^-} - \eta^+ \frac{\partial}{\partial\eta_\mp} \right), \label{n4}
\end{equation}
while the spin part ${W}_{(s)\pm}$ acts on $\tilde{\psi}^{(s_+,s_-)}(k)$ as 
\begin{equation}
	{W}_{(s)\pm}  \tilde{\psi}^{(s_+,s_-)}(k) = 
	\begin{cases} -\frac{i}{\sqrt{2}} \rho s_- \tilde{\psi}^{(s_+,s_- -1)}(k) \label{n5}\\
	              \frac{i}{\sqrt{2}} \rho s_+ \tilde{\psi}^{(s_+-1,s_-)}(k), 
								\end{cases}
\end{equation}
so that ${W}_{(s)\pm}$ lowers the $\mp$ helicity of $\tilde{\psi}^{(s_+,s_-)}(k)$ by one. 

When computing $W_\pm \tilde{\psi}_0(\eta,k)$, (\ref{n4}) will always contribute with a $\eta^+$ term which can be removed by an $\epsilon$ gauge transformation, while (\ref{n5}) will lower the helicity $s_\mp$ by one unit so that we get
\begin{equation}
	W_\pm \tilde{\psi}_0(\eta,k) = \mp i \sqrt{2} \rho \eta_\pm \tilde{\psi}_0(\eta,k) + \text{$\epsilon$ gauge transformation}. \label{n6}
\end{equation}
We then find 
\begin{equation}
	W_+ W_- \tilde{\psi}_0(\eta,k) = 2 \rho^2 \eta_+\eta_- \tilde{\psi}_0(\eta,k)  + \text{$\epsilon$ gauge transformation}. \label{n7}
\end{equation}
Using the notation of this Section the $\eta$-hyperboloid is written as 
\begin{equation}
	2\eta^+\eta^- - 2 \eta_+\eta_- +1 =0, \label{p7}
\end{equation}
so that we  find on the $\eta$-hyperboloid that $W_+ W_- \tilde{\psi}_0(\eta,k) = \rho^2 \tilde{\psi}_0(\eta,k) + \text{$\epsilon$ gauge transformation}$, since the $\eta^+\eta^-$ term in (\ref{p7}) gives rise to a $\epsilon$ gauge transformation. This confirms that the gauge fixed solution has the right eigenvalue for $W^2$ on the $\eta$-hyperboloid. Notice that in Section \ref{ssi1} we showed the same result but with no gauge fixing. 

Finally, we will show that our gauge fixed solution satisfies (\ref{b3}). We could start with the solution for $\psi_0(\eta,x)$ on the $\eta$-hyperboloid, where each helicity appears once, but calculations soon become extremely complicated. We then choose to start in the cotangent bundle, where each helicity appears an infinite number of times, and at the end go back to the $\eta$-hyperboloid. In (\ref{b3}) the states have well defined helicity. Since $\tilde{\psi}_0(\eta,k)$ has zero helicity we must multiply it by $\eta_\pm$ or derive it with respect to $\eta_\pm$ to get non-vanishing helicity and then we must show that when properly normalised they fulfill (\ref{b3}). We find that only products of $\eta_\pm$ and $\tilde{\psi}_0(\eta,k)$ satisfy (\ref{b3}) and are given by 
\begin{eqnarray}
	\chi^{(r,0)}(\eta,k) &=& (\sqrt{2} \eta_+)^r \, \tilde{\psi}_0(\eta,k),  \label{p1} \\
	\chi^{(0,r)}(\eta,k) &=& (\sqrt{2} \eta_-)^r \, \tilde{\psi}_0(\eta,k),  \label{p2}
\end{eqnarray}
where $r$ is a non-negative integer. Notice that (\ref{p1}) and (\ref{p2}) have positive and negative helicity $\pm r$, respectively. If we now use (\ref{n6}) we find that
\begin{eqnarray}
	W_+ \chi^{(r,0)}(\eta,k) &=& - i \rho \chi^{(r+1,0)}(\eta,k) + \text{$\epsilon$ gauge transformation}, \label{p3} \\
	W_- \chi^{(0,r)}(\eta,k) &=&   i \rho \chi^{(0,r+1)}(\eta,k) + \text{$\epsilon$ gauge transformation}, \label{p4} 
\end{eqnarray}
while
\begin{eqnarray}
 W_+ \chi^{(0,r)}(\eta,k) &=& -2i \rho \eta_+\eta_- \chi^{(0,r-1)}(\eta,x) + \text{$\epsilon$ gauge transformation}, \label{p5}\\
 W_- \chi^{(r,0)}(\eta,k) &=&  2i \rho \eta_+\eta_- \chi^{(r-1,0)}(\eta,x) + \text{$\epsilon$ gauge transformation}. \label{p6}
\end{eqnarray}
Then (\ref{p3}) and (\ref{p4}) satisfy (\ref{b3}) on the cotangent bundle while (\ref{p5}) and (\ref{p6}) satisfy (\ref{b3}) only after the use of (\ref{p7}) and absorbing the $\eta^+\eta^-$ term in a gauge transformation, that is, on the $\eta$-hyperboloid. Hence, $\chi^{(r,0)}$ and $\chi^{(0,r)}$ have 
well-defined helicity and satisfy (\ref{b3}) so they describe a CSP with continuous spin $\rho$ on the $\eta$-hyperboloid. 

\section{Wigner Conditions for a CSP \label{d16}}

In \cite{Wigner:1948zp} Wigner found a set of covariant equations for the wave function of one-particle states describing a CSP. In momentum space the wave function $\varphi(\eta,x)$  depends on the momentum $k^\mu$ and on an internal variable $\eta^\mu$ and must satisfy 
\begin{eqnarray}
	 && ik\cdot \partial_\eta \, \varphi(\eta,k) + \rho \, \varphi(\eta,k) = 0, \label{m1} \\
	 && (\eta^2 + 1) \, \varphi(\eta,k)=0, \label{m2}\\
	 && i k\cdot \eta \, \varphi(\eta,k) = 0, \label{m3} \\
	 && k^2 \, \varphi(\eta,k)=0, \label{m4}
\end{eqnarray}
with the last two equations being a consequence of the first two. Our results require that the wave function must be entirely written in terms of $\psi_0(\eta,x)$ and $\psi_1(\eta,x)$ with the $\epsilon$ gauge completely fixed. The gauge choice (\ref{j2}) leads to $\psi_1=0$ so that the wave function depends only on $\psi_0$. The gauge transformation (\ref{d8}) allowed us to gauge away all $+$ components of $\tilde{\psi}_0$ while the $-$ components are expressed in (\ref{j5}) in terms of $\tilde{\psi}_{0 i_1\dots i_n}$. 
We must now recast these conditions on $\tilde{\psi}_0$  in a covariant way in order to find the conditions (\ref{m1})-(\ref{m4}). 

We must start with the full $\tilde{\psi}_0(\eta,k)$ where all components of $\eta^\mu$ are present. The condition that the $+$ components of $\tilde{\psi}_0$ are absent can be implemented as $\delta({\eta^+}) \tilde{\psi}_0$ which can be written in covariant form as $\delta(i k \cdot \eta) \tilde{\psi}_0$ since it reduces to the former expression in the Lorentz frame where $k_- = k_i =0$. Equation (\ref{j5}), which eliminates the {\it minus} components of $\tilde{\psi}_0$, can be enforced as $\Delta \tilde{\psi}_0=0$ 
in the gauge where the $+$ components of $\tilde{\psi}_0$ vanish. We must also recall that the solution of the field equation (\ref{c5}) was extended to all of $\eta$ space 
and that we must go back to $\eta$-hyperboloid by means of a $\delta(\eta^2+1)$. Altogether this means that the wave function must have the form 
\begin{equation}
	\varphi(\eta,k) = \delta(\eta^2+1) \delta(i k \cdot \eta) \tilde{\psi}_0(\eta,k), \label{m8}
\end{equation}
with $k^2 \tilde{\psi}_0 = \Delta \tilde{\psi}_0=0$. It can easily be  checked that (\ref{m8})  satisfies all of the Wigner conditions. This provides an alternative  way to show that (\ref{2.1}) really describes a CSP. 

\section{Final Remarks \label{u1}}

As we have seen the gauge theory for a free CSP is highly non trivial from a mathematical point of view. The use of a cotangent bundle over flat spacetime seems to be the appropriate geometrical setting for its formulation leading to a more complicated framework than that of conventional field theories over Minkowski spacetime. It was shown that the presence of a Dirac delta function and its derivative in the field equation  can be dealt with by solving the delta function constraint, which requires the dynamics to be confined up to the first neighbourhood of the $\eta$-hyperboloid of the cotangent bundle. This leads to a conventional field equation in the cotangent bundle which can be treated by the usual field theory techniques. Then, at the end, we must always return to the $\eta$-hyperboloid or its first neighbourhood as required initially by the delta function constraint. This seems to be the starting point to explore CSPs in a systematic way. A proposal for a gauge theory for fermionic CSPs has been presented \cite{Najafizadeh:2015uxa} and all techniques developed in this paper can be straightforwardly applied to the fermionic case. 

The next step is the introduction of interactions. It is easy to minimally couple the CSP field to an abelian gauge field $A_\mu(x)$. The CSP field $\Psi(\eta,x)$ is now complex and the action reads
\begin{equation}
	S = \frac{1}{2} \int d^4 x \,\, d^4\eta \left[ \delta^\prime(\eta^2+1) D_x \Psi^*(\eta,x) \cdot D_x \Psi(\eta,x) + \frac{1}{2} \delta(\eta^2+1)  \Delta \Psi^*(\eta,x)  \Delta \Psi(\eta,x) \right], \label{u3}
\end{equation}
where $D_x \Psi = (\partial_x - i A)\Psi$, $D_x \Psi^* = (\partial_x + i A)\Psi^*$, $\Delta \Psi = (\partial_\eta \cdot D_x + \rho)\Psi$ and $\Delta \Psi^* = (\partial_\eta \cdot D_x + \rho)\Psi^*$. The action is clearly invariant under the abelian gauge transformation
\begin{equation}
	\delta_\lambda A(x) = \partial_x \lambda(x), \qquad \delta_\lambda \Psi(\eta,x) = i \lambda(x) \Psi(\eta,x), \label{u4}
\end{equation}
and it is also invariant under the $\chi(\eta,x)$ transformation of (\ref{2.2})
\begin{equation}
	\delta_\chi \Psi(\eta,x) = \frac{1}{4} (\eta^2+1)^2 \chi(\eta,x), \qquad \delta_\chi A(x) = 0, \label{u5}
\end{equation}
since the Dirac delta function structure of the action is the same. However, it is not invariant under the CSP $\epsilon$ gauge transformation 
\begin{equation}
	\delta_\epsilon \Psi(\eta,x) =  \eta\cdot D_x \epsilon(\eta,x) - \frac{1}{2} (\eta^2+1) \Delta \epsilon(\eta,x), \label{u6}
\end{equation}
even for a constant abelian gauge field background. This result was expected since the action (\ref{u3}) is regular in the limit $\rho\rightarrow 0$ and it reduces to the action of an infinite number of HS spin fields minimally  coupled to an abelian gauge field which is known for not supporting such interaction\footnote{See \cite{Ponomarev:2016jqk} for a recent discussion and earlier references on interacting HS theory in flat spacetime.}. On the other side we know that to have self-interacting HS particles with spin greater than two in flat spacetime an infinite tower of particles with all spins is required inducing higher derivative interactions which, of course, need dimensionful coupling constants\footnote{See for instance \cite{Vasiliev:2016xui} for earlier work.}. Since we have an action for a CSP with a dimensionful constant $\rho$ it is not at all excluded the existence of self-interacting CSPs with vertices involving higher derivatives. 

As it is well known it is possible to have a quadratic HS theory in (A)dS spaces \cite{Fronsdal:1978vb} and a formulation using a field $\Psi(\eta,x)$ was developed in \cite{Segal:2001qq}. We then expect that a quadratic CSP theory may also be formulated in (A)dS spaces with the limits $\rho\rightarrow 0$ and $\Lambda\rightarrow 0$ being regular.  
We can then wonder whether it would be possible to construct a self interacting CSP theory in (A)dS. Since we have now two free parameters $\rho$ and $\Lambda$ we have much more freedom than in Vasiliev's HS case \cite{Fradkin:1987ks}. It would be very interesting to find the relationship between CSPs and HS fields in (A)dS if the interacting CSP theory do in fact exist.

Another quite important point is that our results with $\rho=0$ are very interesting by themselves since they provide an alternative formulation for an infinite tower of HS fields in flat spacetime. They might shed some light on the old interaction problem of HS fields \cite{Ponomarev:2016jqk}.

\acknowledgments

I would like to thank Xavier Bekaert and Andrei Mikhailov for comments. 
This work was supported by FAPESP Grants 2014/18634-9 and 2011/11973-4. 


%

\end{document}